\documentclass[trackchanges]{aastex631}

\newcommand{\speed}[1]{#1 km~s${}^{-1}$}
\newcommand{\accl}[1]{#1 m~s${}^{-2}$}
\shorttitle{H$\alpha$ Broadening and red asymmetry of a stellar flare}
\shortauthors{Yuchuan Wu et al.}

\graphicspath{{./}{figures/}}

\begin{document}

\title{Broadening and  {redward} asymmetry of H$\alpha$ line profiles observed by LAMOST during a stellar flare  {on an M-type star}}

\author{Yuchuan Wu}
\affiliation{School of Earth and Space Sciences, Peking University, Beijing, 100871, China.}

\author{Hechao Chen}
\affiliation{School of Earth and Space Sciences, Peking University, Beijing, 100871, China.}

\author{Hui Tian}
\affiliation{School of Earth and Space Sciences, Peking University, Beijing, 100871, China.}
\affiliation{National Astronomical Observatories, Chinese Academy of Sciences, Beijing, 100101, China.}


\author{Liyun Zhang}
\affiliation{College of Physics, Guizhou University, Guiyang 550025, China.}
\affiliation{Key Laboratory for the Structure and Evolution of Celestial Objects, Chinese Academy of Sciences, Kunming 650011, China.}

\author{Jianrong Shi}
\affiliation{National Astronomical Observatories, Chinese Academy of Sciences, Beijing, 100101, China.}

\author{Han He}
\affiliation{CAS Key Laboratory of Solar Activity, National Astronomical Observatories, Chinese Academy of Sciences, Beijing 100101, China.}
\affiliation{University of Chinese Academy of Sciences, Beijing 100049, China.}

\author{Hongpeng Lu}
\affiliation{School of Earth and Space Sciences, Peking University, Beijing, 100871, China.}

\author{Yu Xu}
\affiliation{School of Earth and Space Sciences, Peking University, Beijing, 100871, China.}

\author{Haifeng Wang}
\affiliation{GEPI, Observatoire de Paris, Universit\'e PSL, CNRS, Place Jules Janssen 92195, Meudon, France}

\correspondingauthor{Hechao Chen; Hui Tian}
\email{hechao.chen@pku.edu.cn; huitian@pku.edu.cn}

\begin{abstract}

 {Stellar flares are characterized by sudden enhancement of electromagnetic radiation in stellar atmospheres. So far much of our understanding of stellar flares comes from photometric observations, from which plasma motions in flare regions could not be detected.}
From the spectroscopic data of LAMOST DR7, we have found one stellar flare that is characterized by an impulsive increase followed by a gradual decrease in the H$\alpha$ line intensity  {on an M4-type star, and the total energy radiated through H${\alpha}$ is estimated to be on the order of $10^{33}$ erg.} The H$\alpha$ line appears to have a Voigt profile during the flare, which is likely caused by  {Stark} pressure broadening due to the dramatic increase of electron density  {and/or opacity broadening due to the occurrence of strong non-thermal heating}. Obvious enhancement has been identified at the red wing of the H$\alpha$ line profile after the impulsive increase of the H$\alpha$ line intensity. The red wing enhancement corresponds to plasma moving away from the Earth at a velocity of 100$-$200 km s$^{-1}$.   {According to the current knowledge of solar flares, this red wing enhancement may originate from: (1) flare-driven coronal rain, (2) chromospheric condensation, or (3) a filament/prominence eruption that either with a non-radial backward propagation or with strong magnetic suppression.} The total mass of the moving plasma is estimated to be on the order of  {$10^{15}$} kg.

\end{abstract}


\section{Introduction} \label{sec:intro}

Stellar flares are dramatic explosions in stellar atmospheres, which are often characterized by a rapid rise followed by a slow decrease in the electromagnetic radiation across a wide range of wavelength from radio to $\gamma$ rays. On late-type main-sequence stars, these flare events are generally thought to be triggered by sudden release of magnetic energy in the magnetized stellar atmospheres \citep{1991ARA&A..29..275H}. 
According to the widely accepted solar flare scenario \citep[e.g.,][]{2010ARA&A..48..241B,2011LRSP....8....6S}, magnetic reconnection and its associated rapid energy release in the corona play a key role in triggering solar flares. During this process an enormous amount of energy is transported to the lower solar atmosphere and dissipates in the dense chromosphere via thermal conduction and thermalization of (relativistic) energetic electrons. This leads to intense localized plasma heating, which in turn results in chromospheric evaporation and condensation mainly in the impulsive phase of flares \citep[e.g.,][]{2011SSRv..159...19F}. In the meantime, continuous increase of the local temperature, electron density and plasma velocity would cause strong enhancement and broadening of chromospheric spectral lines such as the H$\alpha$ line. Hence, spectroscopic observations of chromospheric lines have long been important diagnostics for the physical processes behind solar and stellar flares.

After decades of multi-wavelength imaging observations of solar flares, solar physicists have well realized that a solar flare is only part of a large-scale eruptive energy releasing process, in which the flare occurs low in the atmosphere and a coronal mass ejection (CME) may take place at a much larger scale \citep[e.g.,][]{2003NewAR..47...53L,2005ARA&A..43..103Z,2011LRSP....8....6S,2011LRSP....8....1C,2017IAUS..328..243O}. In fact, CMEs play the most important role in driving hazardous space weather in the near-Earth space environment \citep{1993JGR....9818937G}. Statistically, flares and CMEs on the Sun are more closely associated with each other with increasing flare class \citep[e.g.,][]{2009IAUS..257..233Y,2011SoPh..268..195A}. Recent photometric observations from the Kepler spacecraft and the Transiting Exoplanet Survey Satellite (TESS) revealed a lot of superflares on G-, K-, and M-type stars {\citep[e.g.,][]{2012Natur.485..478M,2013ApJ...771..127N,2015ApJS..221...18H,2018ApJS..236....7H,2017ApJ...851...91N,2020MNRAS.494.3596D,2020ApJ...890...46T,2021ApJS..253...35T,2021MNRAS.505L..79Y}. The total released energy of one such superflare ($10^{33}-10^{38}$ erg) often far exceeds that of the strongest solar flare (with energy up to $10^{32}$ erg)\citep{2000ApJ...529.1026S,2013PASJ...65...49S}, indicating that they are possibly associated with monster CMEs. Studying stellar flares and associated CMEs of these late-type stars, especially active M dwarfs, are of particular interest. This is because most known potentially habitable exoplanets are orbiting around these stars \citep{2021arXiv211109704A}. If frequent high-energy flares and associated CMEs take place on these stars, they may impact or even threaten the habitability of nearby exoplanets \citep{2007AsBio...7..167K,2008SSRv..139..437Y}. However, detection of CMEs on these stars is still challenging, since direct imaging of stellar coronae is almost impossible even for the nearest stars \citep[see a brief review by][]{2019ApJ...877..105M}.

On the Sun, flare processes can be spatially resolved in great detail. As a result, characteristics of the line broadening and asymmetry during different phases of solar flares have been well studied through many observations \citep[e.g.,][]{1972ARA&A..10....1S,1987ApJ...317..956S,1988ApJ...324..582Z}. For instance, a symmetric chromospheric line broadening is often believed to be caused by the Stark effect or turbulence \citep[e.g.,][]{2017ApJ...837..125K,2019ApJ...879...19Z}. While an asymmetric line broadening is usually explained by bulk plasma motion in the form of chromospheric evaporation \citep[e.g.,][]{1990ApJ...363..318C,2011ApJ...727...98L,2021arXiv211206118L},  {chromospheric condensation \citep[e.g.,][]{1984SoPh...93..105I,1985ApJ...289..425F,1993ApJ...416..886G,1995SoPh..158...81D,2015ApJ...811..139T,2016ApJ...832...65Z,2018PASJ...70..100T,2020ApJ...896..154Y}} or coronal rain \citep[e.g.,][]{2020PPCF...62a4016A,2020ApJ...905...26L,2021arXiv211201354C,2021ApJ...916....5S}.

As an extension of solar spectroscopic studies, spectroscopic observations of chromospheric lines on other stars have not only been used to understand the dynamics of stellar flares, but also been used to detect flare-associated CMEs, since CMEs generally host a chromosphere-temperature dense core, i.e., the erupting cool prominence/filament \citep[e.g.,][]{{1985JGR....90..275I}}. For instance, \citet{1990A&A...238..249H} found an obvious enhancement in the blue wings of Balmer lines during a flare occurring on the well-known flaring dMe dwarf, AD Leo. The blueward asymmetry corresponds to a high speed of \speed{5800}, which has been explained as a possible CME. With spectroscopic observations, \citet{2006A&A...452..987C} conducted a systematic flare study on AD Leo, in which three strongest ones are found to be associated with Balmer line broadening and two show weak redward asymmetries. \citet{2016A&A...590A..11V} reported a complex CME with falling-back and re-ejected material on the M4 dwarf V374 Peg, in which its associated blue shift (\speed{675}) is above the stellar surface escape velocity ($\sim$\speed{580}). From high-resolution spectra of 28 active M dwarfs, \citet{2018A&A...615A..14F} identified broad, potentially asymmetric, wings in H$\alpha$ during stellar flares. They explained the observed red asymmetries as a process similar to chromospheric downward condensation or coronal rain, and suggested that the line broadening may be caused by the Stark effect. 
Recently, several authors have tried to search for CMEs on some flaring M dwarfs with time-resolved spectra \citep{2020MNRAS.499.5047M,2021PASJ...73...44M}. Two candidates of erupting filaments/prominences have been reported on EV Lac and YZ CMi, respectively. With photometric and follow-up spectroscopic observations, \citet{2021ApJ...916...92W}  {reported detection of} possible CMEs during flares of two M-dwarfs.
In addition, focusing on more than 5500 spectra of M-dwarf stars, \citet{2019A&A...623A..49V} found 478 spectra with Balmer-line asymmetries, most of which are associated with flux enhancement and their maximum velocities are on the order of 100$-$\speed{300}. Similarly, \citet{2020A&A...637A..13M} and \citet{2021A&A...646A..34K} respectively used a large number of spectra to search for flares and CMEs on M dwarfs, but consistently reported a high level of flare activity and a low detection rate of CME candidates.  {More recently, \citet{2021NatAs.tmp..246N} successfully detected a definite evidence of a flare-filament eruption event on the G-type star EK Draconis using time-resolved H${\alpha}$ spectroscopic observations; moreover, they compared this stellar eruption signature with solar filament eruptions in the Sun-as-a-star spectral view, revealing that the former is the same as the magnified version of the latter.}

From the spectra provided by the Large Sky Area Multi-Object Fibre Spectroscopic Telescope (LAMOST), we have found one interesting stellar flare with obviously broadened and asymmetric H$\alpha$ line profiles. In Section 2, we describe the observational data and show analysis results of the event, including the Voigt-Gauss function fitting and red-blue asymmetry analysis. In Section 3, we estimate the flare energy and mass of the moving plasma, and provide explanations for the Lorentz profile and asymmetry of H$\alpha$. Finally, conclusions are given in Section 4.

\section{Observation and data analysis results} \label{sec:style}


LAMOST, also called the Guoshoujing Telescope, is a medium-aperture ( {4.6} m) optical telescope in China, with the capability of performing spectral surveys of the northern-sky stars in the Milky Way and other objects \citep{1998SPIE.3352...76S,2012RAA....12.1197C}. 
In recent years, LAMOST provides spectra in time domain, making it possible to search for stellar flares through the time evolution of H$\alpha$ line intensity and profile asymmetry. 
Here we used data from LAMOST DR7 to search for H$\alpha$ line asymmetries. In LAMOST DR7 medium-resolution (R$\sim$7500) spectral catalogue, there are 2 828 655 time-domain spectra, covering a wavelength range of 6300$-$6800 \AA. These spectra were taken between 2017 September 28 and 2019 June 8. 

\begin{figure}[ht!]
\centering {\includegraphics[width=18cm]{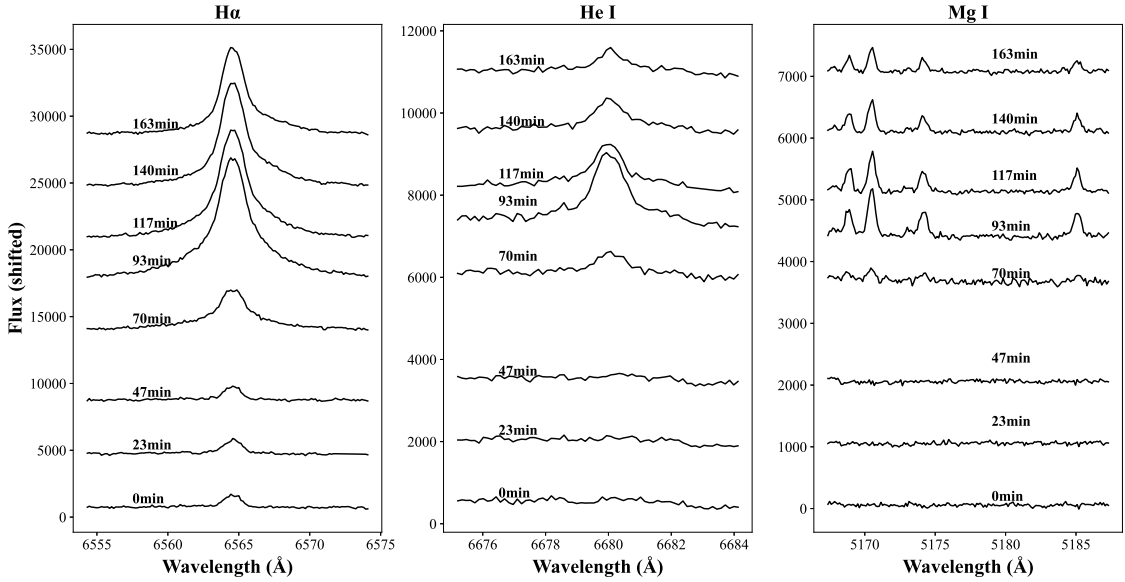}} 
\caption{H$\alpha$ {, He{~\sc{i}} and Mg{~\sc{i}}} line profiles observed at different times during the flare of interest. The profiles are shifted in the y-axis for a better illustration. \label{fig:general}}
\end{figure}

The stellar flare we study here was observed on 2019 January 16, and the starting observation time is 23:37 (local time)  {, or 15:37 (UT)}. The obsid is 716907097. The original time-domain spectra of the H$\alpha$ {, He{~\sc{i}} 6680 \AA~and Mg{~\sc{i}} 5171 \AA~}lines are shown in Figure 1. Each exposure lasts for 20 minutes, and the time cadence of exposures is around 23 minutes. From Figure 1, we can see that obvious radiation enhancement and profile broadening  {of H$\alpha$ line} commence at 70 mins since the starting time and last to the end of this observation.  {Other chromospheric lines like He{~\sc{i}} and Mg{~\sc{i}} become visible during the flare, and their intensities decrease in the later phase of the flare.} Interestingly, a red wing enhancement of the H$\alpha$ line was also  {presented} in the later phase of the flare, indicating the presence of materials (plasma) moving  {away from} the Earth.  {To get a more clear view of the red wing enhancement, please see Figure 2 in Section 2.1.}  {LAMOST provides Signal-to-noise ratios (SNRs) that were calculated by the equation of $flux * (inverse \ variance) ^{0.5}$. For the eight spectra of the star, their SNRs are 10.02, 9.96, 10.13, 27.04, 23.64, 13.27, 11.62 and 10.13, respectively. Note that after taking the Earth's motion and the star's redshift into consideration, the corrected theoretical line center of H${\alpha}$ profile is 6564.61 \AA.}

According to LAMOST catalog  {\footnote{\url{http://dr7.lamost.org/v2.0/spectrum/view?obsid=432616084}}}, this target is an M4-type star,  {and other relevant data are shown in Table 1.}  {The star was also observed by the Kepler's second mission \citep{2014PASP..126..398H}, and several flares were detected on it \footnote{{The K2 observations of this star can be found from \url{https://mast.stsci.edu/portal/Mashup/Clients/Mast/Portal.html} or \url{https://drive.google.com/drive/folders/1K46MbuEarPNwNwRIr0n8dR7eCppkYzDy?usp=sharing}}}. We used the effective temperature to estimate the mass and radius of the star. We can find the average radius, mass and effective temperature of an M2 or M5 type main sequence star from \citet{2000PhT....53j..77C}. We then assumed a linear relationship of $lg(\frac{M}{M_{\sun}})$ and $lg(\frac{R}{R_{\sun}})$ versus $lgT_{\rm eff}$ for stars with a spectral type between M2 and M5. Here $M$, $M_{\sun}$, $R$ and $R_{\sun}$ represent the masses and radii of the star and our Sun, respectively, and $T_{\rm eff}$ is the effective temperature of a star. Since the effective temperature of the observed star is 3404 K, we obtained the following,
\begin{equation}
\frac{M}{M_{\sun}}=0.326, \frac{R}{R_{\sun}}=0.411.
\end{equation}
Then we can calculate the escape velocity of the star, which is \speed{550}.

\begin{deluxetable*}{ccccccccc}
\tablenum{1}
\tablecaption{ {Basic information of the star (716907097)}}
\tablewidth{0pt}
\tablehead{
\colhead{Right ascension} & \colhead{Declination} & \colhead{Subclass} & \colhead{Redshift} & \colhead{Effective temperature} & \colhead{Surface gravity} & \colhead{Magnitudes} & \colhead{Magnitudes} & \colhead{Magnitudes} \\
\colhead{(degree)} & \colhead{(degree)} & \colhead{} & \colhead{} & \colhead{(K)} & \colhead{(log $g/cm~s^{-2}$)} & \colhead{of g filter} & \colhead{of r filter} & \colhead{of i filter}
}
\startdata
127.52 & 16.57 & dM4 & 0.00012 & 3404 & 5.02 & 15.82 & 14.53 & 13.04 \\
\enddata
\end{deluxetable*}

\subsection{Normalized spectra} \label{subsec:tables}

The continuum levels of the original H$\alpha$ spectra vary with time, e.g., in the fourth and fifth exposures the continuum levels are about twice as large as in other exposures. This variation is too big to be explained by a white-light flare. Instead, it is more likely caused by the variation of seeing. Due to the lack of absolute flux calibration, we thus assumed that the continuum level remains unchanged before, during, and after the flare.  {However, we also point out that the H$\alpha$ flare might be associated with a white-light flare, especially considering that the broadened H${\alpha}$ profiles observed in the flare (see Figure 2) indicate a non-thermal heating process \citep[see details in Section 3.3 and also][]{2020PASJ...72...68N}, but we cannot confirm whether a white-light flare exists from the available data.} Before further analysis, we need to normalize the original spectra to the continuum intensities. To obtain the continuum intensities, we used a summation of a Voigt function, a Gauss function (only for the last four spectra) and a constant background to fit each H$\alpha$ profile. The constant was taken as the continuum level. The Gauss function was introduced here to account for the additional bulk flow component, which can be inferred from the redward asymmetry. The Voigt function is a convolution of a Lorentz function and a Gauss function. A Voigt function fitting has often been applied to analyze the spectral broadening of hydrogen Balmer lines (i.e., H${\alpha}$, H$_{\beta}$) \citep[e.g.,][]{2003JPhB...36.1573L,2009ApJ...696.1755T,2015ApJ...806..214F} and helium lines \citep[e.g.,][]{2021ApJ...916....5S,2021arXiv211114647K} in astrophysical and laboratory plasmas. Here, we used it to fit the H$\alpha$ line profile because the chromospheric line broadening is likely dominated by a combination of pressure broadening and Doppler broadening during flares, which can be described as a Voigt function \citep{2019ApJ...879...19Z}. The total fitting function can be written as follows \citep{2011ascl.soft09001G},

\begin{equation}
I(\lambda)=\frac{Re[wofz(\frac{\lambda-\lambda_{cen}+i\gamma}{\sqrt{2}\sigma})]}{\sqrt{2\pi}\sigma}I_{0}+ae^{-(\frac{\lambda-m}{s})^{2}}+C,  
\end{equation}

where $\lambda$ and $\lambda_{cen}$ are the wavelength and centroid wavelength of the H$\alpha$ line, respectively. $Re$ represents the real part of a complex number. And $wofz$ is the Faddeeva function, which satisfies $wofz(z)=e^{-z^{2}}erfc(-iz)$.

There are 7 free parameters: $I_{0}$ characterizes the total flux of the Voigt function; $\gamma$ is the half width at half maximum (HWHM) of the Lorentzian component. A larger $\gamma$ means that the Voigt function is more similar to a Lorentz profile, indicating a larger effect of pressure broadening. The parameter $\sigma$ is the standard deviation of the the Guassian component in the Voigt function. The height, centroid and width of the Gauss function are represented by $a$, $m$ and $s$, respectively. The continuum level is represented by $C$. By dividing the original spectrum taken during each exposure by $C$, we obtained the normalized spectrum $I_{n}(\lambda)$. All normalized spectra are shown in Figure 2. 

\begin{figure}[ht!]
\centering {\includegraphics[width=14cm]{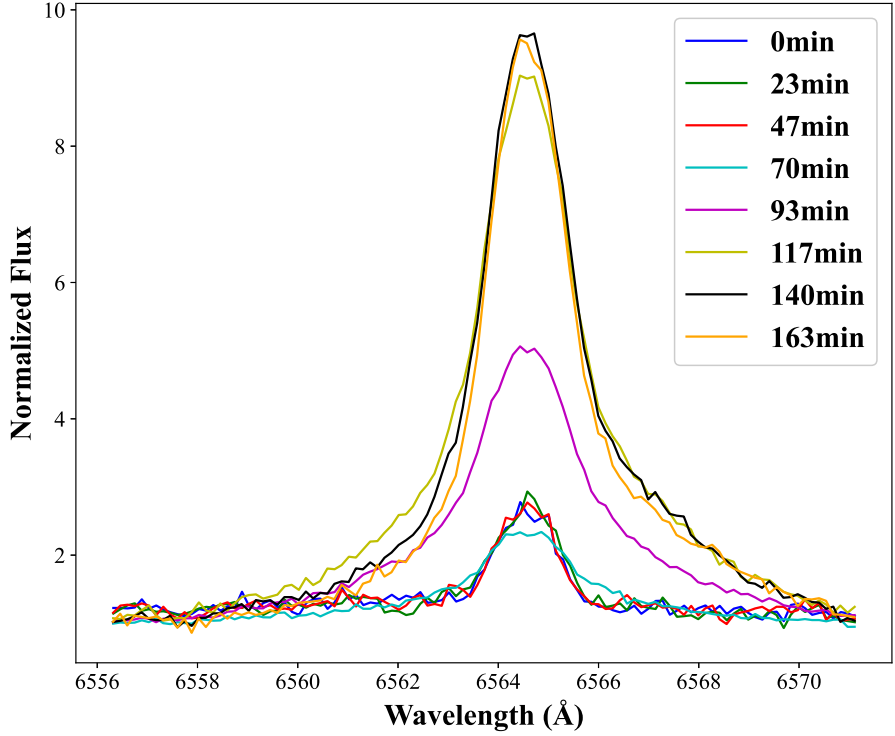}} 
\caption{H$\alpha$ line profiles normalized to the continuum intensities.  \label{fig:general}}
\end{figure}

\subsection{Voigt-Gauss function fitting} \label{subsec:tables}

We used the same function to fit the normalized spectra. After normalization, the fitting function becomes the following one,

\begin{equation}
I_{n}(\lambda)=\frac{Re[wofz(\frac{\lambda-\lambda_{cen}+i\gamma}{\sqrt{2}\sigma})]}{\sqrt{2\pi}\sigma}I_{n,0}+a_{n}e^{-(\frac{\lambda-m}{s})^{2}}+1.
\end{equation}

The subscript $n$ means normalized intensity. The normalized spectrum, fitting function, and the residual between fitting and observation for each exposure are shown in Figure 3. Note that we removed the continuum just for the sake of simplicity. The variations of $I_{0}$, widths of the Gaussian and Lorentzian components of the Voigt-Gaussian fitting function, as well as the fitting parameters of the Gaussian function are shown in Figure 4.

\begin{figure}[ht!]
\centering {\includegraphics[width=20cm]{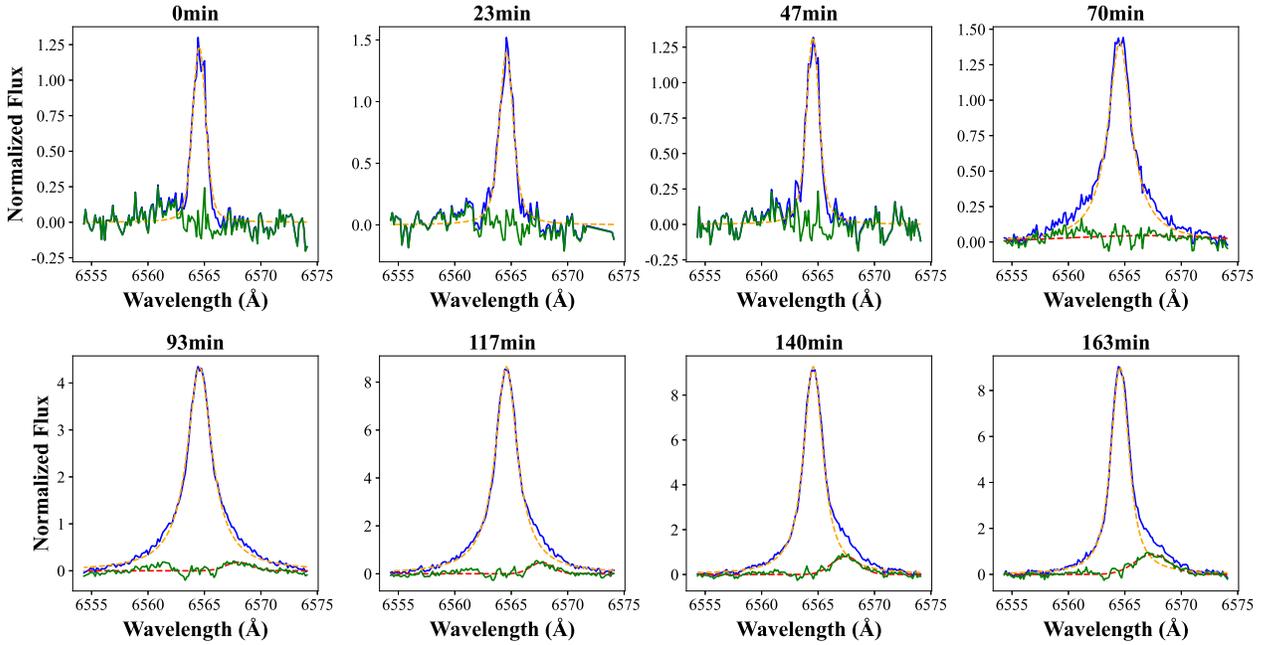}} 
\caption{Normalized H$\alpha$ line profiles and their fitting results. Blue lines are the original data, orange lines are fitting results  {for Voigt function, red lines are fitting results for Gauss function}, and green lines indicate residuals between fitting  {results for Voigt function} and observation. \label{fig:general}}
\end{figure}

\begin{figure}[ht!]
\centering {\includegraphics[width=18cm]{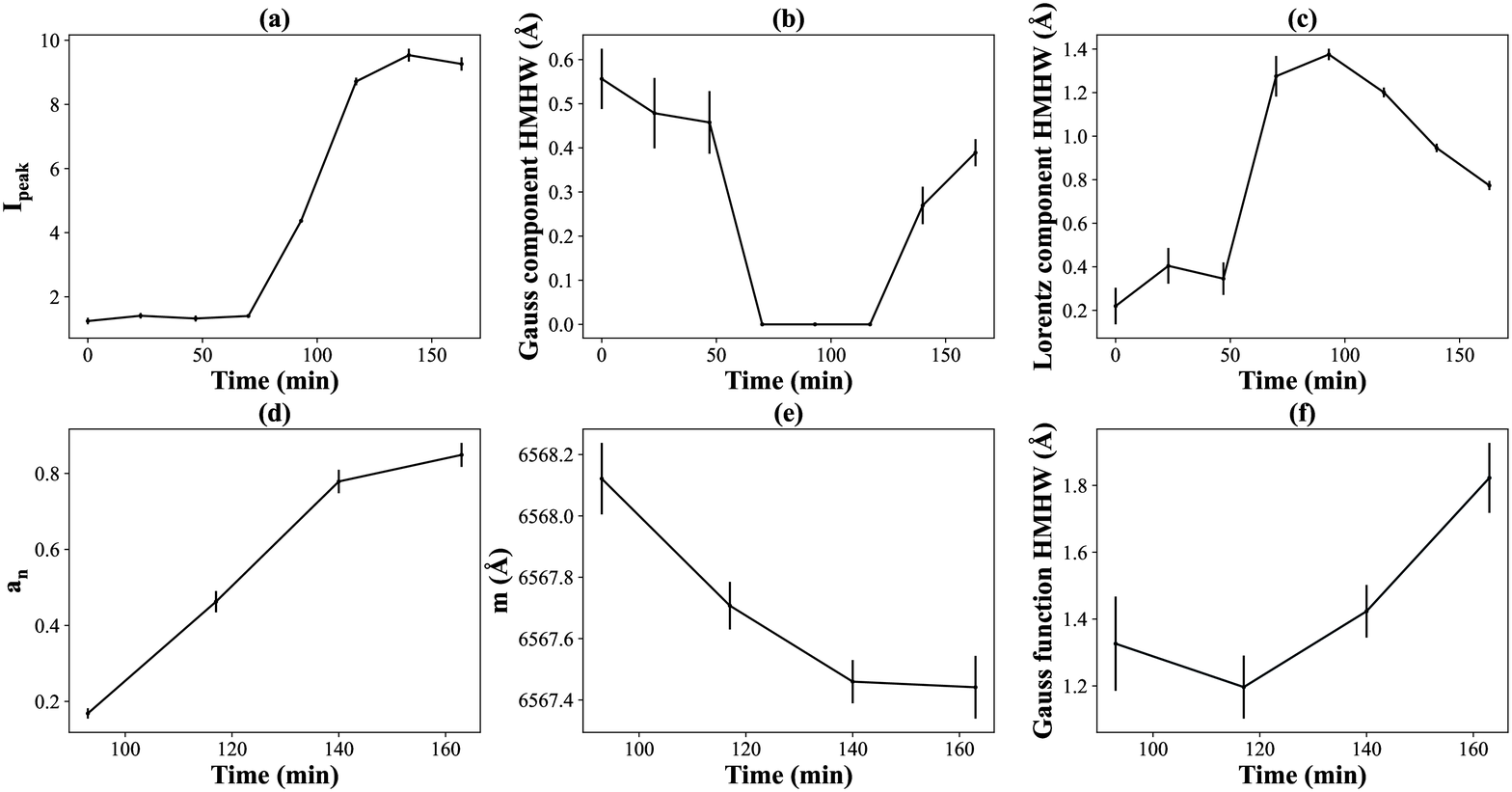}} 
\caption{Time evolution of the fitting parameters. The top rows show the total flux of the Voigt function ($I_{0}$), the Half width at half maximum (HWHM) of the Gaussian component ($\sigma\sqrt{ln4}$), and the HWHM of the Lorentzian component ($\gamma$) in the Voigt-Gauss function, respectively. The bottom row only shows the height ($a_n$), the centroid ($m$), and the HWHM ($s\sqrt{ln2}$) in the Gaussian function for the last four spectra, where profile asymmetries are observed. In each panel, the error bars represent uncertainties calculated from the fitting errors.
\label{fig:general}}
\end{figure}

From Figures 3 and 4, we noticed the following characteristics:

(1) The total flux of the Voigt function, $I_{0}$, increases quickly between 70 min and 117 min, indicating the occurrence of a flare. 

(2) The HWHM of Gaussian component in the Voigt-Gauss fitting function is close to 0 for three spectra taken at 70 min, 93 min and 117 min. At these occasions, the wings of the H$\alpha$ line profiles are clearly characterized as a Lorentz function. Interestingly, this change of spectral profile starts as soon as the rapid increase of $I_{0}$. When the rapid increase of $I_{0}$ stops, the HMHW of Gaussian component in the Voigt-Gauss fitting function begins to increase.

(3) $a_n$ keeps on increasing after 93 min, meaning that the redward asymmetry of the H$\alpha$ line profile becomes more and more significant.

(4) $m$ slowly decreases after 93 min, indicating that the moving plasma responsible for the redward asymmetry of H$\alpha$ possibly decelerates after 93 min (see Figure 3 (e)).  {From 93 min to 163 min, the average deceleration rate was estimated as \accl{7.6}. }

The time variation of the integrated H$\alpha$ flux can reflect the change of radiation power through the H$\alpha$ line. We first integrated each normalized spectrum from 6557 \AA \ to 6572 \AA, then remove the continuum. The remaining integral  {is the equivalent width} of the H$\alpha$ line. Using the fitting parameters of the Gauss function, we can also obtain  {the equivalent width of the redshifted enhancement} responsible for the line asymmetry in the last four exposures. The time variations of  {the equivalent widths of the H$\alpha$ line and the redshifted enhancement} are shown in Figure 5. We can see that the  {the equivalent width of H$\alpha$ line} increases quickly between 70 min and 117 min, and decreases slowly after 117 min, which validates the conclusion from the analysis of the total flux of the Voigt function. At the same time,  {the equivalent width of the redshifted enhancement} simultaneously grows by a factor of $\sim$6 from 93 min to 163 min.

\begin{figure}[ht!]
\centering {\includegraphics[width=10cm]{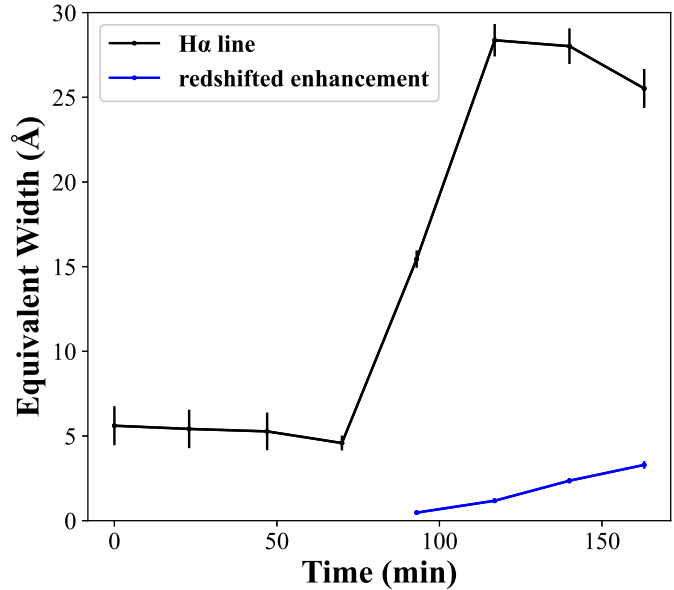}} 
\caption{ Time variations of  {the equivalent widths of the H$\alpha$ line and the redshifted enhancement}. The black error bars indicate uncertainties of  {the equivalent width of the H$\alpha$ line} by summing the measurement errors between 6557 -- 6572 \AA. The blue error bars represent the uncertainties of  {the equivalent width of the redshifted enhancement} as calculated from the fitting errors.
\label{fig:general}}
\end{figure}

\subsection{ Red-blue asymmetry analysis} \label{subsec:tables}

A technique called red-blue (RB) asymmetry analysis has been often used to analyze the asymmetries of solar spectral lines both under the quiet-Sun condition \citep[e.g.,][]{2009ApJ...701L...1D,2011ApJ...738...18T} and during solar eruptions \citep[]{2012ApJ...748..106T}. The technique can reveal where the asymmetry is most obvious. Here we used this method to analyze the asymmetry of the H$\alpha$ line profiles. The RB asymmetry for an offset velocity can be expressed as following:

\begin{equation}
RB(u_{c})=\int_{u_{c}-\delta u/2}^{u_{c}+\delta u/2}I(u)du-\int_{-u_{c}-\delta u/2}^{-u_{c}+\delta u/2}I(u)du,
\end{equation}
where $u_{c}$, $\delta u$, and $I(u)$ respectively represent the velocity from the line center, the velocity range over which the RB asymmetry is determined, and the ratio between spectral intensity and peak intensity. Here the H$\alpha$ line center and peak intensity are $\lambda_{cen}$ and $I_{0}$, respectively, as calculated in Section 2.1. We found that a smaller $\delta u$ leads to too much fluctuation, and a bigger $\delta u$ ignores too much data in the spectra. By trial and error, we found that 0.5 \AA\ is just appropriate for our calculation. So we let $\delta u$ be the velocity corresponding to  0.5 \AA \ Doppler shift, which is around \speed{22.8}. We calculated $RB$ as a function of $u_{c}$, where $u_{c}=\speed{22.8\times(0.5+n)$}, $n=0,1,2,\cdots$.  

The results of the RB asymmetry analysis for the spectra taken during the last four exposures are shown in Figure 6. We can see that an obvious redward asymmetry, which is indicated by the positive peak of an RB asymmetry profile, is clearly present after 93 min, when the H$\alpha$ line intensity shows an impulsive increase. The redward asymmetry becomes more evident towards the end of the observation. The peaks of the RB asymmetry profiles indicate a redshifted velocity of \speed{100} to \speed{200}, far below the escape velocity of the star ({\speed{550}}).

\begin{figure}[ht!]
\centering {\includegraphics[width=10cm]{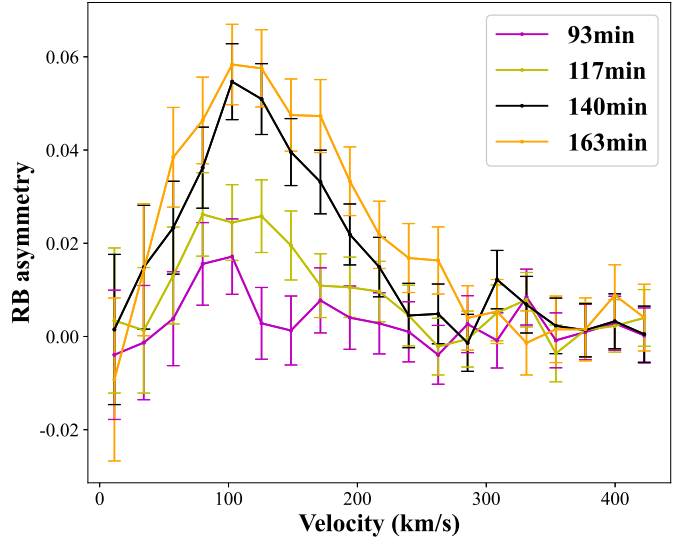}} 
\caption{RB asymmetry analysis results for the H$\alpha$ spectral profiles taken during the last four exposures. The error bars indicate the uncertainties propagated from the measurement errors.
\label{fig:general}}
\end{figure}

\section{Discussion} \label{sec:style}

\subsection{Radiation power  {and total energy} of the flare} \label{subsec:tables}
We tried to estimate the radiation power of the flare. We first estimated the luminosity of the continuum ($L_c$) under the H$\alpha$ line. Since the white light continuum of a star generally can be approximated by the Planck function, $L_c$ can be calculated through the following equation:

\begin{equation}
L_{c}=4\pi^2 R^{2}\int_{6557 \rm \AA}^{6572 \rm \AA}\frac{1}{e^{\frac{hc}{\lambda kT}}-1}\frac{2hc^{2}}{\lambda^{5}}d\lambda= {7.58\times10^{28}}\ erg\ s^{-1},
\end{equation}
where $T$ represents the effective temperature of the star, which is 3404.72 K. $L_{c}$ represents the luminosity of the continuum between 6557 \AA \ and 6572 \AA. $R$ is the radius of the star, which is $0.411R_{\sun}$.

Considering that  {the equivalent width of the normalized continuum} in the wavelength range of 6557$-$6572 \AA~is 15  {\AA \ }and  {the equivalent width} of the H$\alpha$ line has a peak of  {28.36 \AA}, we can calculate the peak luminosity ($L_{f}$) of the flare through the following relation,

\begin{equation}
\frac{L_{f}}{L_{c}}=\frac{ {28.36}}{15}.
\end{equation}

And $L_{f}$ was estimated to be  {$1.43\times10^{29}$} erg $s^{-1}$.

 {We can also estimate the total energy of the flare from the integrated flux calculated in Section 2.2. This observation did not cover the whole process of the flare. We need to remove the H$\alpha$ contribution from non-flare state of the star. We used the average of first three spectra to estimate it, and the result in the form of equivalent width is 5.43 \AA. During the observation, from 70 min to 163 min, the flare energy radiated through H$\alpha$ is $4.45\times10^{32}$ erg. If we assume that the power of the flare and the equivalent width of H$\alpha$ in the last three spectra decay at the same speed of the last three spectra, we estimated that the flare lasts for 418 min, and the total flare energy radiated through H$\alpha$ was estimated to be $2.42\times10^{33}$ erg.}

\subsection{ {Mass of the moving plasma}} \label{subsec:tables}

 {We have tried to estimate the mass of the moving plasma. We decided to use the method which  was first introduced by \citet{1990A&A...238..249H} and recently used by \citet{2021A&A...646A..34K} to estimate the mass.} We used the following formula to obtain a lower limit of the mass \citep{1990A&A...238..249H},

\begin{equation}
 {M\ge \frac{4\pi d^{2}F_{m}(N_{tot}/N_{j})m_{H}\eta_{OD}}{h\upsilon_{j-i}A_{j-i}}=\frac{L_{m}(N_{tot}/N_{j})m_{H}\eta_{OD}}{h\upsilon_{j-i}A_{j-i}},}
\end{equation}
 {where $d$ is the distance from the star, $F_{m}$ is the integrated excess flux of the moving plasma. Since we did not have absolute flux calibration, we used $L_{m}$, the luminosity of the moving plasma, to estimate $4\pi d^{2}F_{m}$.} $N_{tot}/N_{j}$ is the ratio of the number density of hydrogen atoms and the number density of hydrogen atoms at excitation level $j$. $m_{\rm H}$ is the mass of a hydrogen atom. $\eta_{OD}$ is the opacity damping factor. $h$ is the Plank constant. $\upsilon_{j-i}$ and $A_{j-i}$ are the frequency and Einstein coefficient of the spectral line, respectively.

 {As for other parameters in the formula above, we used the same settings as \citet{2021A&A...646A..34K} once used.} This flare event under study was detected from the H$\alpha$ line (transition from j = 3 to i = 2). However,  {they} could not find an appropriate calculation to estimate $N_{tot}/N_{3}$, so  {they} decided to use the relevant data of the H$\gamma$ line (j = 5, i = 2) for  {their} calculation. \citet{1990A&A...238..249H} performed non-local thermal equilibrium (NLTE) modeling for AD Leo and estimated $N_{tot}/N_{5}=2\times10^{9}$.  {They} set $\eta_{OD}$ to 2, which indicates that half of the radiation escapes the moving feature \citep{2014MNRAS.443..898L}. For H$\gamma$, $A_{j-i}$ is $2.53\times10^{6}$ \citep{2009JPCRD..38.1129W}.

We used the same method mentioned in Section 3.1 to estimate the radiation power of the moving plasma through the red wing enhancement of the H$\alpha$ line $L_{m,\alpha}$. In Section 2.3, we have  {calculated the equivalent width of the redshifted enhancement}, and the  {equivalent width} reaches the maximum in the last exposure. The  {equivalent width} is  {3.29} \AA, which means $\frac{L_{m,\alpha}}{L_{c}}=\frac{ {3.29}}{15}$. Through this way we found $L_{m,\alpha}= {1.66}\times10^{28}$ erg s$^{-1}$. To obtain the radiation power of the H$\gamma$ line $L_{m,\gamma}$, we need to divide $L_{m,\alpha}$ by a factor of $\sim$3 \citep{1988A&A...206L...1B}, {just as \citet{2021A&A...646A..34K} did}. Finally, we obtained a lower limit for the mass of the moving plasma, which is  {$3.2\times10^{15}$} kg.
\subsection{Lorentz profile of the H$\alpha$ line} \label{subsec:tables}

From 70 min to 117 min, the H$\alpha$ line appears to reveal a Lorentz profile
 { with a symmetric line broadening. Similar line broadening phenomena have been reported in solar flare observations \citep[e.g.,][]{2017ApJ...837..125K,2019ApJ...879...19Z} and also in stellar flare studies \citep{2018A&A...615A..14F,2020PASJ...72...68N}, which have been interpreted as non-thermal broadening or Stark (pressure) broadening.} There are two classical theories for the explanation of Stark broadening: the impact theory \citep{1932ZPhy...75..287W} and the statistical theory. The former results in a Lorentz profile and the latter leads to a Holtsmark profile \citep{1919AnP...363..577H}. Since the observed H${\alpha}$ spectra can be well fitted by a Voigt-Gauss function that includes a Lorentz profile, here we take the former to explain our current observation. According to the impact theory, each time the radiating atom collides with a charged particle, the radiation is interrupted. After Fourier transform, these interrupted radiation form a Lorentz profile in frequency domain.
 {Moreover, the RADYN numerical simulation of M-dwarf flares from \citet{2020PASJ...72...68N} recently suggested that due to the non-LTE formation properties and highly variable opacity of the H${\alpha}$ line, the broadened H${\alpha}$ profile observed during flares is determined not only by the Stark broadening but also sometimes relates to ``opacity broadening". Indeed, through changing the energy spectra from soft- into hard-energy input cases, their simulation results revealed that once hard high-energy electron beams appear during flares and deposit their energy in the deep chromosphere, strong self-absorptions and a Stark effect can simultaneously cause a wider line broadening. Therefore, this radiative-transfer-induced broadening process may also be important for the broadened flaring H${\alpha}$ line, because stellar and solar flares often associate with intense non-thermal radiation/heating.}

Clear Lorentz profiles were observed at 70 min, 93 min and 117 min, which correspond to the impulsive increase of the H$\alpha$ line intensity. On the basis of the theory mentioned above, pressure broadening implies that the collision between electrons and hydrogen atoms becomes more frequent, that is to say, electron density increases significantly. It is well known that in the impulsive phase of solar flares, energetic electrons are accelerated in the corona, then propagate downward along flare loops and finally collide with hydrogen atoms in the (deep) chromosphere.  {Meanwhile, as revealed in \citep{2020PASJ...72...68N}, if a strong non-thermal heating occurs in this stellar flare, the forming height of H${\alpha}$ line can further extend into the lower chromosphere and then a larger opacity (or strong self-absorptions) would also contribute to a wider line broadening.}

After 117 min, the H$\alpha$ line intensity stops increasing, indicating the end of the impulsive phase. From Fig. 4 we can see that the HWHM of the Gaussian component in the Voigt-Gauss fitting function becomes non-zero again, meaning that the spectral line profiles start to deviate from a Lorentz function and the electron density begins to decrease {, but the spectral line profiles are still wider than their pre-flare state. This likely indicates that a non-thermal heating, i.e., energetic electron beams, is still continuing but with a softer/weaker energy injection. Indeed, \citet{2020PASJ...72...68N} recently reported a good correspondence between the white-light emission and the H${\alpha}$ line width variation in stellar flares of the M-type star AD Leo, suggesting H${\alpha}$ line broadening has a fairly good relationship with non-thermal heating. In their observed flares, a similar decease in the H${\alpha}$ line width from the impulsive to decay phase was also found.} Such a scenario is in accord with the physical explanation of the flare decaying phase in the standard solar flare scenario \citep[e.g.,][]{2010ARA&A..48..241B,2011LRSP....8....6S}.

\subsection{Redward asymmetry of the H$\alpha$ line profile} \label{subsec:tables}

On the Sun, numerous spectroscopic observations of flares have revealed that chromospheric lines often exhibit line asymmetries during flares. Redward asymmetries indicative of velocities of several tens of \speed{} have been frequently reported during the flare impulsive phase. This is commonly thought to be indued by chromospheric condensation \citep[e.g.,][]{1985ApJ...289..425F}, the downward movement of the chromospheric plasma caused by the high pressure in the chromosphere. Downflows of dense plasma draining back to the solar surface along post-flare loops \deleted{may} also lead to redward asymmetries of chromospheric spectral lines \citep[e.g.,][]{2015ApJ...811..139T}. Blue asymmetries or blue shifts of spectral lines have also been detected in the early phase of solar flares, which indicate upflows with speeds of a few to a few hundreds \speed{} in different spectral lines. This has often been attributed to the chromospheric evaporation, namely warm/hot plasma upflows driven by intense chromospheric heating. In the meantime, CMEs may also produce absorption line asymmetries if erupting filament materials are contained in CMEs. For example, \citet{1993ARep...37...76D} reported a strong mass ejection during a solar flare through obvious blue wing absorption in H${\alpha}$, indicating a radial velocity of over \speed{600}. Similarly, \citet{2003ApJ...598..683D} investigated a series of filament-induced absorption in the H${\alpha}$ blue wing, revealing a good temporal correspondence between the acceleration process of a filament and the impulsive phase of a two-ribbon solar flare.

During flares on other stars, Doppler shifts or asymmetries of spectral lines have also been frequently detected \citep[e.g.,][]{1990A&A...238..249H,1994A&A...285..489G,2006A&A...452..987C,2016A&A...590A..11V,2018A&A...615A..14F,2019MNRAS.482..988C}. Using X-ray observations, \citet{2019NatAs...3..742A} recently reported a long-lasting blueshift of $\sim$ \speed{90} of cool plasma (3 MK) at the decay phase of a large stellar flare on an evolved giant star HR 9024. They explained this cool upflow as a CME candidate. At optical wavelengths, spectral line asymmetries are more frequently reported  {\citep[e.g.,][]{2014MNRAS.443..898L,2018PASJ...70...62H,2021A&A...646A..34K}}, although their explantations sometimes are not straightforward.  {For instance, \citet{2018PASJ...70...62H} found that the H${\alpha}$ line in a stellar flare on the dMe star EV Lac exhibited a long-lived blueward asymmetry, with a velocity of \speed{$\sim$ 100}, even during the whole duration of the flare. They speculated that the blue asymmetries might be caused by possible filament activation/eruption and evaporation, but little is known about the definite origin of such long-duration blue asymmetries yet.}
\citet{2021A&A...646A..34K} have searched asymmetries of the Balmer lines in time-resolved spectroscopy provided by the Sloan Digital Sky Survey data release 14 and reported six CME candidates with excess flux in line wings, with the highest projected velocities of \speed{300$-$700}. In fact, most (5/6) of these asymmetries appear at the red wings. \citet{2021A&A...646A..34K} suggested that these red-wing asymmetries may be caused by several possible processes, i.e., backward-directed CMEs occurring close to the limb, falling cool material during the eruption, or severe chromospheric condensation.

In our case, apart from obvious  {line} broadening, red-wing asymmetries also appear during the flare process. From Figure 5, it is clear that the redward asymmetry becomes more evident after the impulsive increase in the H${\alpha}$ intensity. It strongly suggests that such line asymmetries are a flare-induced phenomenon. The line asymmetries we detected last towards the end of our observations, when the H${\alpha}$ intensity starts to decrease. 
 {This is timed to coincide with the line broadening after the flare peak, namely the continuing non-thermal heating. Moreover, the line asymmetries demonstrate an increasing amplitude until the end of our observations, with a redshifted velocity at \speed{100$-$200}.
According to the current knowledge of solar flares, the long-lived red wing enhancement may originate from three possible scenarios: (1) a flare-driven coronal rain process. The typical downward velocity of solar coronal rain is \speed{more than 100}; as revealed in many solar flare observations, coronal rain are resulted from the radiative cool process of hot flaring plasma that trapped in post-flare loops, thus they often become more prominent in the decay phase of flares. These observed features are basically consistent with the properties of red-wing asymmetries we detected here. (2) a flare-induced chromospheric condensation process. In solar flares, chromospheric condensation results from flare non-thermal heating processes and it usually happens in the impulsive phase, with a typical downflow velocity of \speed{tens of} \citep[e.g.,][]{1995SoPh..158...81D,2015ApJ...811..139T,2020ApJ...896..154Y} but sometimes even  up to more than \speed{100} as well \citep [e.g.,][]{1984SoPh...93..105I,2018PASJ...70..100T}. Here, the red-wing asymmetry is accompanied by obvious line broadening after the flare peak, thus suggesting a continuing non-thermal heating might still be responsible for the appearance of condensation downflows in the decay phase. (3) a flare-associated filament/prominence eruption either with a non-radial backward propagation or with strong magnetic suppression. In case of a backward ejected filament, the flare must occur near or at the limb of the stellar disk where its footpoint-produced H${\alpha}$ emission (especially line broadening) at least can be visible from the Earth. This special spatial location requirement suggests that a backward filament eruption at least is a very quite rare case. 
Another situation is a failed filament eruption, which is first reported on the Sun \citep{2003ApJ...595L.135J} and recently supposed to be common on M-type stars due to their strong magnetic suppression \citep{2018ApJ...862...93A,2020ApJ...900..128L}. 
In this case, the ejected filament material may be quickly confined by overlying magnetic field after the flare peak and then demonstrate more and more prominent downward mass falling to the stellar disk, which thus might cause the increasing amplitude of red-wing asymmetries in our observation. {Meanwhile, the ejecting filament at the flare onset is expected to cause a short-lived blueshift before we detected the red-wing asymmetries. But this blueshift was absent in the LAMOST observation, possibly due to its poor temporal resolution (around 23 mins)}. Caution that only using chromospheric spectral observations, our current observation is very difficult to distinguish between these three possible scenarios or their combinations, since the observed H${\alpha}$ radiation from the star could be a superposition of flare, post-flare loop, and prominence eruption. In future, to better understand these physical processes in stellar flares, new time-resolved spectroscopic observations using multi-temperature lines are needed.
} 

The speed of the moving plasma is \speed{100} to \speed{200}, far below the star’s escape velocity. However, we should remember that this speed only represents the lower limit of the true velocity of the moving material. Because the Doppler shift can only be used to measure the velocity component along the line of sight, and ejected filament material originating from the limb very likely has a small velocity component in the line of sight.  {Thus, we can not exclude the possibility that the backward filament eruption might drive a CME if the red-wing asymmetries were to be an eruption. Moreover, as shown in Figure 4 (e), the moving plasma responsible for the redward asymmetry likely decelerates. The average deceleration rate is computed as \accl{7.6}, which is much smaller than the gravity of the star. In the filament/prominence eruption scenario, this very small deceleration rate might supports that the ejecting filament material is flying along a non-radial direction that is near perpendicularly to the line of sight, suffering a severe projection effect.}

 {On the other hand,} the acceleration of plasma during solar coronal rain is about 1/3 $g_{\rm eff}$ \citep{2020PPCF...62a4016A}. If we assume that the falling materials move downward with a constance acceleration of 1/3 $g_{\rm eff}$ and the propagation distance scales with the radius of the star, the maximum velocity of the falling plasma on a star can be estimated as follows,
\begin{equation}
v=v_{\sun}\sqrt{\frac{gR}{g_{\sun}R_{\sun}}},
\end{equation}
where $v_{\sun}$ is the maximum velocity of plasma during coronal rain on the Sun, which is about \speed{100} according to \citet{2020PPCF...62a4016A}. $g$ and $g_{\sun}$ are the gravitational accelerations on the star and the Sun, respectively. And the stellar and solar radii are represented by $R$ and $R_{sun}$, respectively. Our calculation yielded a value of {\speed{126}} for $v$, which is consistent with the velocities inferred from our RB asymmetry analysis,  {supporting the possible coronal rain scenario mentioned above.} 

\section{Conclusion} \label{sec:style}

From the LAMOST DR7 data archive, we identified a strong flare event on an M4-type star. The key findings are summarized as follows,

(1) The flare shows an impulsive increase followed by a gradual decrease in the H$\alpha$ line intensity. The impulsive increase starts at 70 min and the H$\alpha$ intensity peaks at 117 min since the beginning of the observation. The maximum radiation power of the flare through the H$\alpha$ line is  {$1.43\times10^{29}$} erg s$^{-1}$.  {During the observation, the flare radiated energy in H$\alpha$ is $4.45\times10^{32}$ erg, and the total flare energy radiated through H$\alpha$ is estimated to be on the order of $10^{33}$ erg.}

(2) During the impulsive phase, the H$\alpha$ line appears to have a Lorentz profile, which we think is caused by downward propagating nonthermal electrons colliding with hydrogen atoms in the chromosphere. After t=117 min, the spectral profile gradually deviates from a Lorentz profile due to the  {decrease} of non-thermal electron acceleration and the  {resultant} decrease of electron density in the chromosphere.

(3) H$\alpha$ line asymmetries corresponding to plasma moving away from the Earth at a velocity of \speed{100$-$200} are detected. The asymmetries might  {result from three possible scenarios: (1)} flare-driven coronal rain,  {(2) chromospheric condensation, or (3) a filament/prominence eruption either with a non-radial backward propagation or with strong magnetic suppression.} The total mass of the moving plasma is estimated to be on the order of  {$10^{15}$} kg.

\section{Acknowledgements} \label{sec:style}

H.H. is supported by the National Key R\&D Program of China (2019YFA0405000).
This work is also supported by NSFC grants 11825301, 11790304,  \& 11973059, the fellowship of China National Postdoctoral Program for Innovative Talents (BX20200013), and Peking University Undergraduate Research Program.  {We thank the referee for the very helpful comments and constructive suggestions, and} thank Y.-J. Zhu, Dr. Y. Li, and  Dr. H. Li} for helpful discussions. 
The Large Sky Area Multi-object Fiber Spectroscopic Telescope, LAMOST (Guoshoujing Telescope) is a National Major Scientific Project built by the Chinese Academy of Sciences. Funding for the project has been provided by the National Development and Reform Commission. LAMOST is operated and managed by the National Astronomical Observatories, Chinese Academy of Sciences.

\end{document}